\documentclass[12pt]{elsarticle}
\usepackage{amsmath}
\usepackage{graphics}
\usepackage{graphicx} 
\usepackage{amsfonts}
\usepackage{setspace}
\usepackage{subfigure}

\begin{document}
\begin{frontmatter}

\title{Two-Dimensional Modeling of Ideal Merging Plasma Jets}
\author{J. Loverich}
\author{A. Hakim}
\tnotetext[*]{This research was supported in part by the Department of Energy SBIR
Grant DE-SC0000833. The fluids code uses the FACETS
framework\cite{facets} and the authors would like to thank the FACETS
framework team, J Cary, A Hakim, A Pletzer, S Shasharina, M Miah
(Tech-X Corporation) and T Epperly (Lawrence Livermore National
Laboratory).}
\address{Tech-X Corporation \\ 5621 Arapahoe Avenue, Suite A, \\ Boulder CO, 80303}
\begin{abstract}
Idealized merging argon plasma jets are simulated in 2D using both gas dynamic and MHD models.  Results indicate
that peak pressures of several hundred kilobar can be achieved for high Mach number jets.  Including
a simple optically thin Brehmstrahlung radiation model and plasma targets shows that extremely high densities and magnetic fields
can be achieved during jet merging on the order of ~1000 times the initial density/field.  Further investigations should
include detailed ionization processes and more accurate radiation modeling to properly capture the radiation transport
and subsequent target compression.
\end{abstract}
\end{frontmatter}

\section{Introduction}
Recent interest in magneto inertial fusion\cite{Thio1999, Lindemuth1995, Degnan1999} has resulted in the investigation of new schemes for magnetized target compression.  The typical
foil liner approach suffers from the ``stand off" problem where mechanical connections to the device are destroyed with each shot.  Many different
solutions to the stand off problem have been described.  One possible approach is the
use of plasma jets for target compression\cite{Thio2001,Thio1999,Hsu2008,Parks2008,Cassibry2008}.   In this scenario the target is moved into a chamber and a cylindrical array of 
plasma jets is fired simultaneously at the target.  The jets merge to form a plasma liner that compresses the target.  A new experiment at Los Alamos National Lab, the Plasma Liner Experiment (PLX) is
currently under construction to explore plasma jet merging.  This work is part of an effort to understand jet merging physics and develop numerical tools for modeling such
a device.  In particular this is a first effort at investigating jet merging with radiation. 

The current investigation uses ideal MHD for magnetized target simulations and the ideal Euler equations for simulations without magnetic fields.   The MHD model used is single 
species so both the target and the jet are made of argon, in an ideal experiment the target plasma would be hydrogenic so a multi-species MHD model would be used for modeling.  In the MHD case a simple optically thin Bremsstrahlung radiation model assuming constant ionization state is used.    The effects of simple radiation model are so dramatic that it becomes clear that precise knowledge of plasma chemistry and radiation transport will be very important in future theoretical/numerical investigations.  Ultimately a collisional radiative model combined with a radiation diffusion is likely needed to make accurate predictions.

The results presented show typical peak pressure and densities that could be achieved in a high powered merging jet experiment in the ideal case.  In an actual experiment with similar jet 
parameters where jets are merging in 3D, peak pressures are likely
to be significantly higher.  The results also illustrate ``radiative collapse" which could occur in this type of device if the jets were to remain optically thin during liner 
formation and collapse.  The loss of thermal energy due to radiation in the jet during 
compression might mean that the ram efficiency of the jet remains high during the target compression phase resulting in efficient coupling of jet directed energy to target compression. 
\section{Model}
Both gas dynamic \eqref{E:GasDynamic} and magnetohydrodynamic (MHD) \eqref{E:MHD} models are used in this paper.  The gas dynamic model is equivalent to the MHD model when the magnetic field is set to zero everywhere, however the gas dynamic model is used because it is  faster computationally since there only 5 unknowns instead of 8.  The gas dynamic problem is used in situations where there is no magnetic field.  An ideal equation of
state is assumed and a simple radiation model is used in several simulations.  The radiation loss term is given in \eqref{E:Brehms} 
as
\begin{equation}\label{E:Brehms}
	P_{\text{Br}} = n_{e}\,n_{i}\left(1/7.69\times 10^{18}\right)^{2}T_{e}^{1/2}Z^{2}_{\text{eff}}
\end{equation}
where the temperature $T_{e}$ is given in electron volts, $n_{e}$ is the electron number density in $\text{\#}/m^{3}$ and $Z_{\text{eff}}$ is the effective charge state.
\begin{equation}\label{E:GasDynamic}
\frac{\partial}{\partial t}\left(
  \begin{array}{c}
  \rho \\
  \rho\,u_{x} \\
  \rho\,u_{y} \\
  \rho\,u_{z} \\
  e \\
  \end{array}
  \right) +
  \nabla\cdot\left(
  \begin{array}{ccc}
  \rho\,u_{x} & \rho\,u_{y} & \rho\,u_{z} \\
  \rho\,u_{x}^{2}+P & \rho\,u_{x}\,u_{y} & \rho\,u_{x}\,u_{z} \\
  \rho\,u_{y}\,u_{x} & \rho\,u_{y}\,u_{y} + P & \rho\,u_{y}\,u_{z} \\
  \rho\,u_{z}\,u_{x} & \rho\,u_{z}\,u_{y} & \rho\,u_{z}\,u_{z} + P \\
  u_{x}\left(e+P\right) & u_{y}\left(e+P\right) & u_{z}\left(e+P\right)\\
  \end{array}
  \right) = 
  \left(\begin{array}{c}
  0 \\
  0 \\
  0 \\
  0 \\
  -P_{\text{Br}}\\
  \end{array}\right)
\end{equation}
Ideal MHD equations with a bremsstrahlung radiation loss term are used for problems with a magnetized target
\small
\begin{equation}\label{E:MHD}
\frac{\partial}{\partial t}\left(
  \begin{array}{c}
  \rho \\
  \rho\,u_{x} \\
  \rho\,u_{y} \\
  \rho\,u_{z} \\
  e \\
  B_{x} \\
  B_{y} \\
  B_{z} \\
  \Psi \\
  \end{array}
  \right) +
  \frac{\partial}{\partial x}\left( \begin{array}{c}
  \rho\,u_{x}  \\
  \rho\,u_{x}^{2}+P - \frac{1}{2\,\mu_{0}}B_{x}^{2} + \frac{1}{\mu_{0}}B^{2} \\
  \rho\,u_{y}\,u_{x}-\frac{1}{\mu_{0}}B_{x}\,B_{y}  \\
  \rho\,u_{z}\,u_{x}-\frac{1}{\mu_{0}}B_{x}\,B_{z}  \\
  u_{x}\left(e+P+\frac{1}{2\mu_{0}}B^{2}\right)-\frac{1}{\mu_{0}}B_{x}\,B\cdot u   \\
  \Psi  \\
  \left(u_{x}\,B_{y}-u_{y}\,B_{x}\right)  \\
  -\left(u_{x}\,B_{z}-u_{z}\,B_{x}\right) \\
  \Gamma^{2}\,B_{x}  \\
  \end{array}\right)+\frac{\partial F_{y}}{\partial y}+\frac{\partial F_{z}}{\partial z} = 
    \left(\begin{array}{c}
  0 \\
  0 \\
  0 \\
  0 \\
  -P_{\text{Br}}\\
  0 \\
  0 \\
  0 \\
  0 \\
  \end{array}\right)
\end{equation}
\normalsize
The variable $\Psi$ is the correction potential.  This term is used to propagate errors in the constraint equation $\nabla\cdot B=0$ out of the domain at the error propagation speed $\Gamma$.  This
technique was originally developed in \cite{Dedner2002}.
The electron number density $n_{e}$ is computed from these equations as $n_{e}=Z_{\text{eff}}\frac{\rho}{m_{i}}$ where $m_{i}$ is the ion mass and the 
temperature is given by $T_{e}\left(\text{kelvin}\right)=\frac{P}{k_{b}\,n\left(Z_{\text{eff}}+1\right)}$.  The simulations performed in this paper are 2D meaning that the Z derivative is zero, however all 
terms including $u_{z}$, $B_{x}$ and $B_{y}$ are kept even though their values remain zero throughout the simulation.  Nautilus is a 3D code, however, 2D simulations were performed so that peak
pressure convergence could be achieved.  At low resolution the predicted peak pressure is dominated by numerical diffusion and is much lower than it should be.  At a resolution of 800X800 cells or higher the
numerical diffusion no longer dominates the measured peak pressure and convergence is achieved.  This also suggests that in 3D, converged simulations would 
require 800X800X800 cells and would take significantly longer to run.
\section{Numerical Approach}
Simulations were performed using Nautilus, a fluid plasma modeling code developed at Tech-X.  In order to obtain jet propagation
through vacuum a specialized approach was required.  The approach taken is related to that used in
\cite{Waagan2009, Stone2009} and is described below.

In these simulations the equations are solved at every time step using two different approaches.  The first approach
is a positivity conserving first order finite volume method.  This technique maintains positive pressure and positive density in the solution even
when there are density or pressure jumps of 9-12 orders of magnitude.  The scheme is locally conservative and shock capturing, though only
first order accurate.  A first order positive solution throughout the entire domain is computed using this technique.  This solution in cell $i$ is 
called $q^{p}_{i}$ and the solution has associated with it one flux across each face $s$ denoted $f^{p}_{s}$.  In this case $s$ can be thought of
as a global index of cell faces since each face is shared by two cells.

To compute a higher order accurate solution a second scheme is used.  The second scheme is a second order upwind MUSCL method which is also shock
capturing and conservative, but not positivity conserving.  The solution obtained using this approach in cell $i$ is denoted $q^{a}_{i}$ and has associated with it one flux across each face $s$ denoted $f^{a}_{s}$.

Solving just using the first order accurate scheme results in solutions that are far too diffuse at the resolutions used in this paper and the jets expand much more rapidly than they should.  As a result, it's necessary to use a combination of the first and second order scheme.

Every face has associated with it one flux, the flux is added to one cell and subtracted from the other, this ensures conservation.  The final solution is computed from a combination of fluxes from the positive solution and the accurate solution.  The final flux $f_{s}$ chosen for face $s$ becomes

\begin{equation}
	f_{s} = \left[
	\begin{array}{c}
	f^{a}_{s}\text{ if } P^{a}_{R} > P_{0} \text{ and } \rho^{a}_{R} > \rho_{0} \text{ and } P^{a}_{L} > P_{0} \text{ and } \rho^{a}_{L} > \rho_{0}\\
	f^{p}_{s}\text{ otherwise}
	\end{array}\right]		
\end{equation}
where $P^{a}_{R}$, $\rho^{a}_{R}$, $P^{a}_{L}$ and $\rho^{a}_{L}$ are the pressure and densities determined by the accurate scheme 
to the left and the right of face $s$.  $P_{0}$ and $\rho_{0}$ are basement densities
and pressures that are typically set to $1\times 10^{-6}$ to $1\times 10^{-9}$ times the peak initial pressure and density in
 the simulation.  This approach ensures that the positive solution is used in cells where the pressure and density are predicted to be 
 below the threshold.  
 
 It's important to note, that if one simply sets a basement pressure and density (as is frequently done in MHD), mass, 
 momentum and energy conservation errors will occur that can drastically alter the results to these jet problems introducing unphysical 
 effects that destroy the solution. This is especially true when radiation losses are included.  In our specific case, simply setting the background
 pressure and density to basement values introduced perturbations in the solution which caused small shock waves in our high speed
 jets and they lost symmetry before merging even occurred.  This effect became exaggerated when the jets moved diagonal to the grid.  Finally,
 if the background density and pressure were too high, say 1/1000 the jet density and pressure, the jets would create a large wake which would
 then interact with neighboring jets resulting in the development of shocks in the jet even before jet merging.
\section{Setup}
Parameter regimes for jet merging were explored by Cassibry\cite{Cassibry2009} and by Parks\cite{Parks2008}.  Based on
those results we've chosen jets that are in the fusion reactor regime.  The PLX device is a spherical device where the jets are placed at a subset of the faces
of a truncated icosahedron with up to 32 jets.  In these 2D simulations we've used 6, 12 and 24 jets to gain some idea of the relationship between peak
pressure and the number of jets in non-radiating simulations.  In these gas dynamic simulation no target is included.  The initial temperatures, jet diameters and pulse lengths are based on simulations
suggested by Cassibry\cite{Cassibry2009} for the PLX experiment. In the simulations that follow the following conditions are used.  Jets are initialized to either
Mach 80 = 226 km/s or Mach 40 = 113 km/s.  The 2D jet diameter is 8cm with a jet length of 50cm.  The initial total temperature
is 2eV while the initial number density is assumed to be $5.0\times10^{23} \frac{1}{m^{3}}$.  The plasma is an argon plasma and the bremsstrahlung
radiation model assumes constant ionization state Z=9.  Densities in the vacuum are set to $1\times 10^{-6}$ times the initial jet density and vacuum pressures are set to $1\times 10^{-8}$ times the initial jet pressure.  The results have been split into two sets, the first set of results are gas-dynamic simulations where we are simply looking at the peak pressure and density ignoring radiation effects.  These simulations help us bound the parameter regime for simulations with plasma chemistry.  In the second set of simulations we use one jet configuration and look at the effect of our simple radiation model with a magnetized target.  In the radiating case we've only included one jet configuration as that is sufficient to illustrate the important physical effect observed, namely the high peak density and magnetic field due to radiation losses in the merging jets and the target.  Modifying the number of jets in the radiating case did not change the observed phenomena.

\section{Results}
\begin{figure}
	\begin{center}
		\includegraphics[scale=0.3]{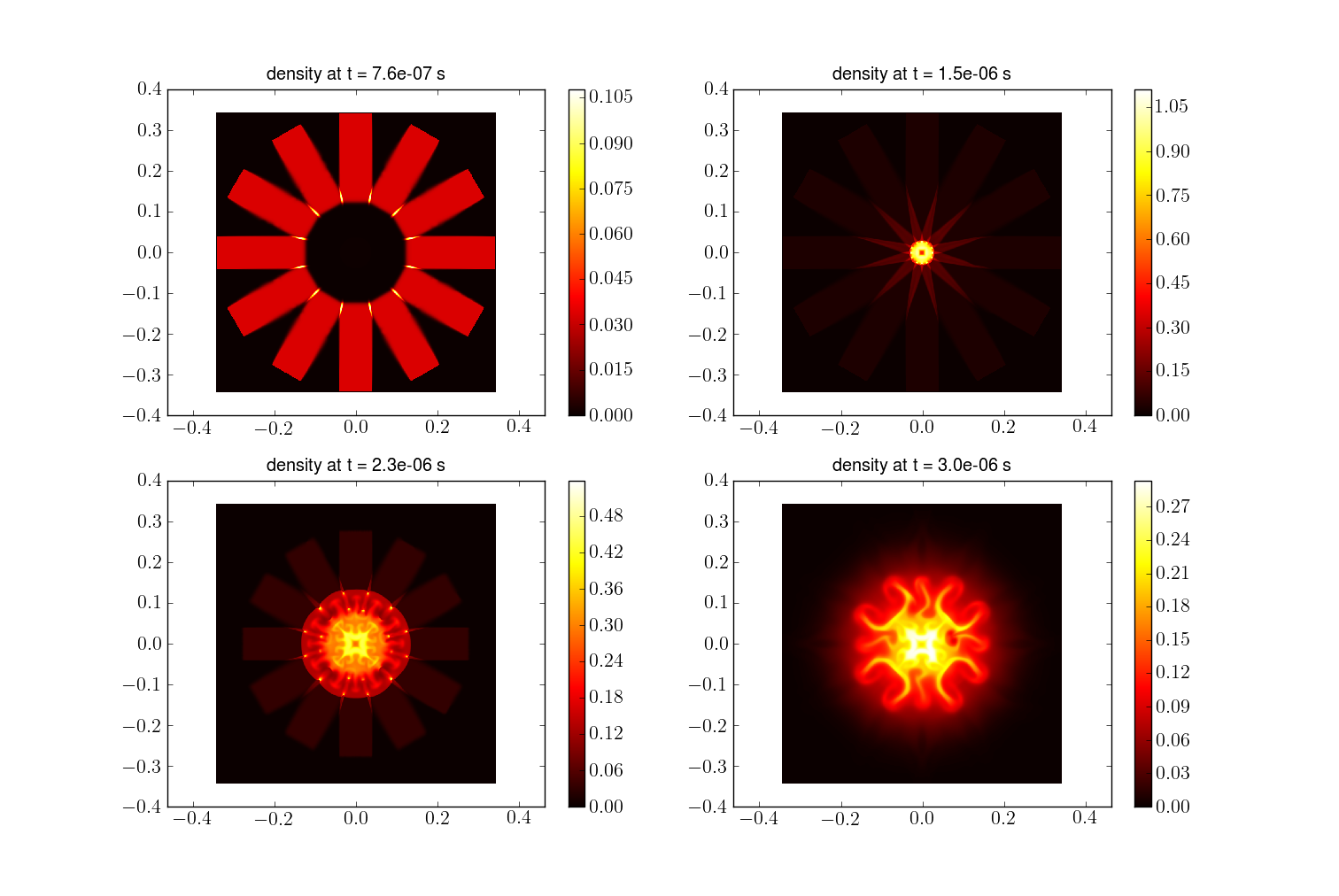}
	\end{center}\caption{Simple jet merging using inviscid Navier Stokes equations.  Snapshots of mass density.  Simulation 
	is 2D and we are looking down on plane of the jets.  The first image shows liner formation as the jets begin to merge.  Second image shows
	a series of shockwaves as the jets interact at the center.  Third image shows solution as the shockwave expands outwards.  4th image
	shows the development of Rayleigh Taylor instabilities.}\label{F:densityPlot}
\end{figure}

\begin{figure}
	\begin{center}
		\includegraphics[scale=0.3]{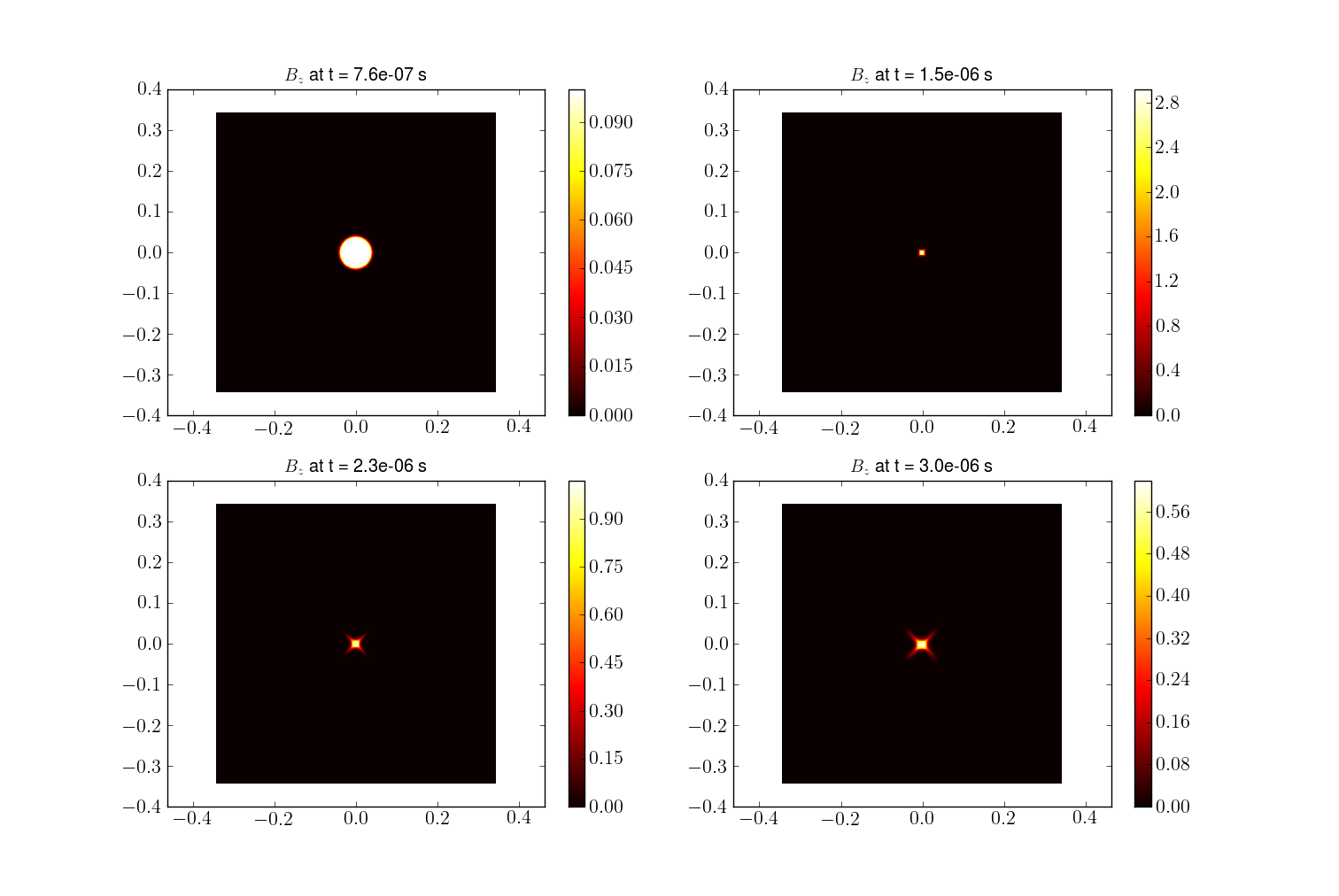}
	\end{center}\caption{Simple jet merging using ideal MHD equations with a magnetized argon target (no radiation).  
	Snapshots of $B_{z}$.  The first image shows the magnetic field in the target before the jets have begun to interact with the target field.  The second image shows
	the target near peak compression from the jets.  The third image shows expansion of the target and the fourth image shows the development of Rayleigh Taylor instabilities. }\label{F:mach80Bz}
\end{figure}

Figure \ref{F:densityPlot} shows results of a neutral fluid simulation at Mach 80 on a $1600\times1600$ grid.  Four snapshots are shown.  The first  sub plot (upper left) shows the jets just as they begin to merge and the liner is formed.  Shocks develop where the jets begin to merge and are indicated by bright lines in the plot.  The second sub plot (upper right) shows the jets after they've collided in the middle and the resulting shock wave begins expanding outwards.  The third sub plot (lower left) shows additional shock expansion.  Final sub plot (lower right) shows the development of Rayleigh Taylor instabilities.
Figure \ref{F:mach80Bz} shows the same simulation using MHD with a $0.1$ Tesla magnetized target.  The first sub plot (upper left) shows the initial $B_{z}$ field.  The second plot (upper right) shows the field near peak compression.  The next two plots show the field again as it expands.  Some Rayleigh Taylor instability is evident.

\begin{figure}
	\begin{center}
		\includegraphics[scale=0.4]{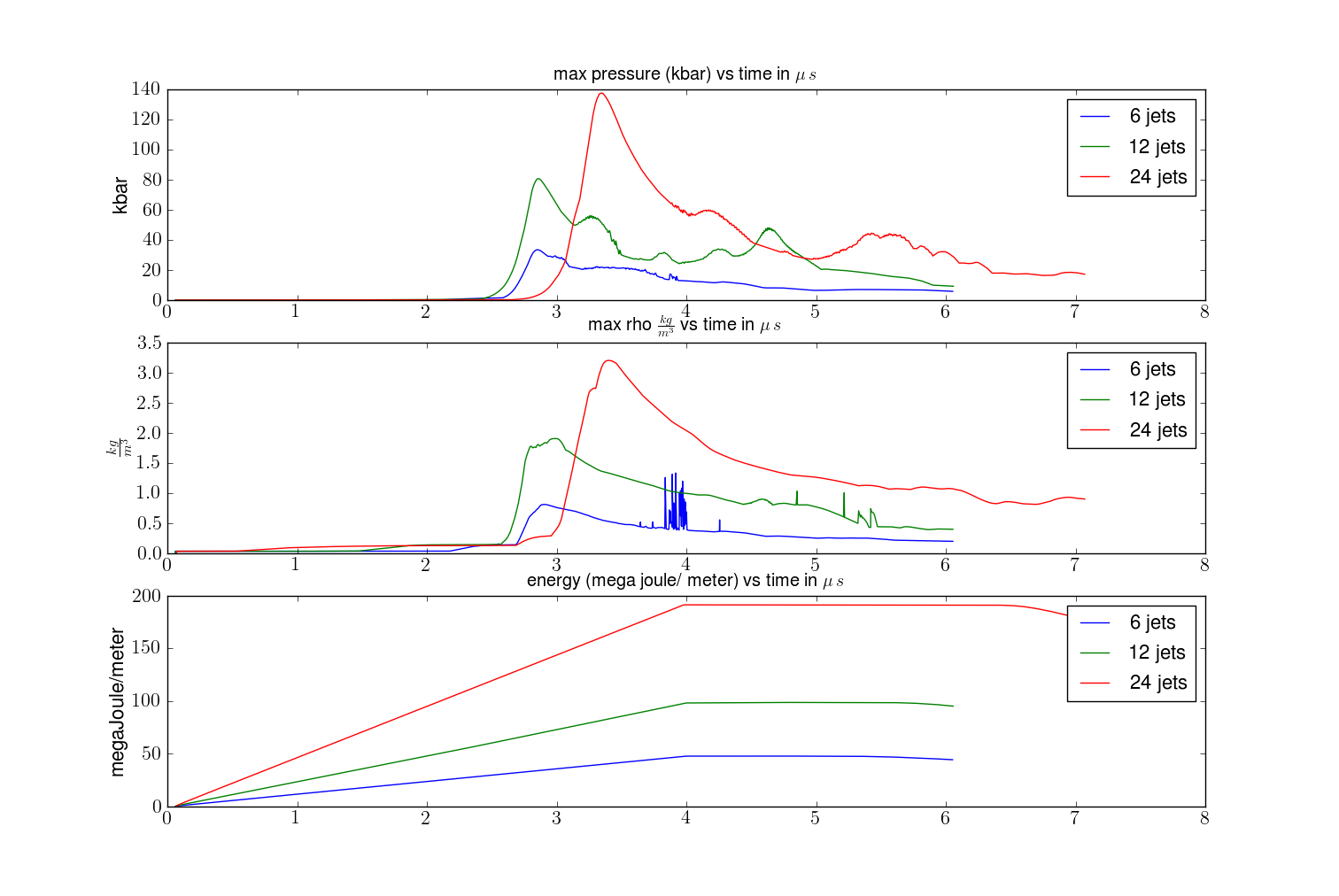}
	\end{center}\caption{Merging Mach 40 gas dynamic jets without radiation.  Total energy is recorded as $MJ/meter$ since the simulation 
	is 2D.  Peak pressure reaches $140 \text{kbar}$ in the 24 jet simulation.}\label{F:eulerHist40}
\end{figure}

\begin{figure}
	\begin{center}
		\includegraphics[scale=0.4]{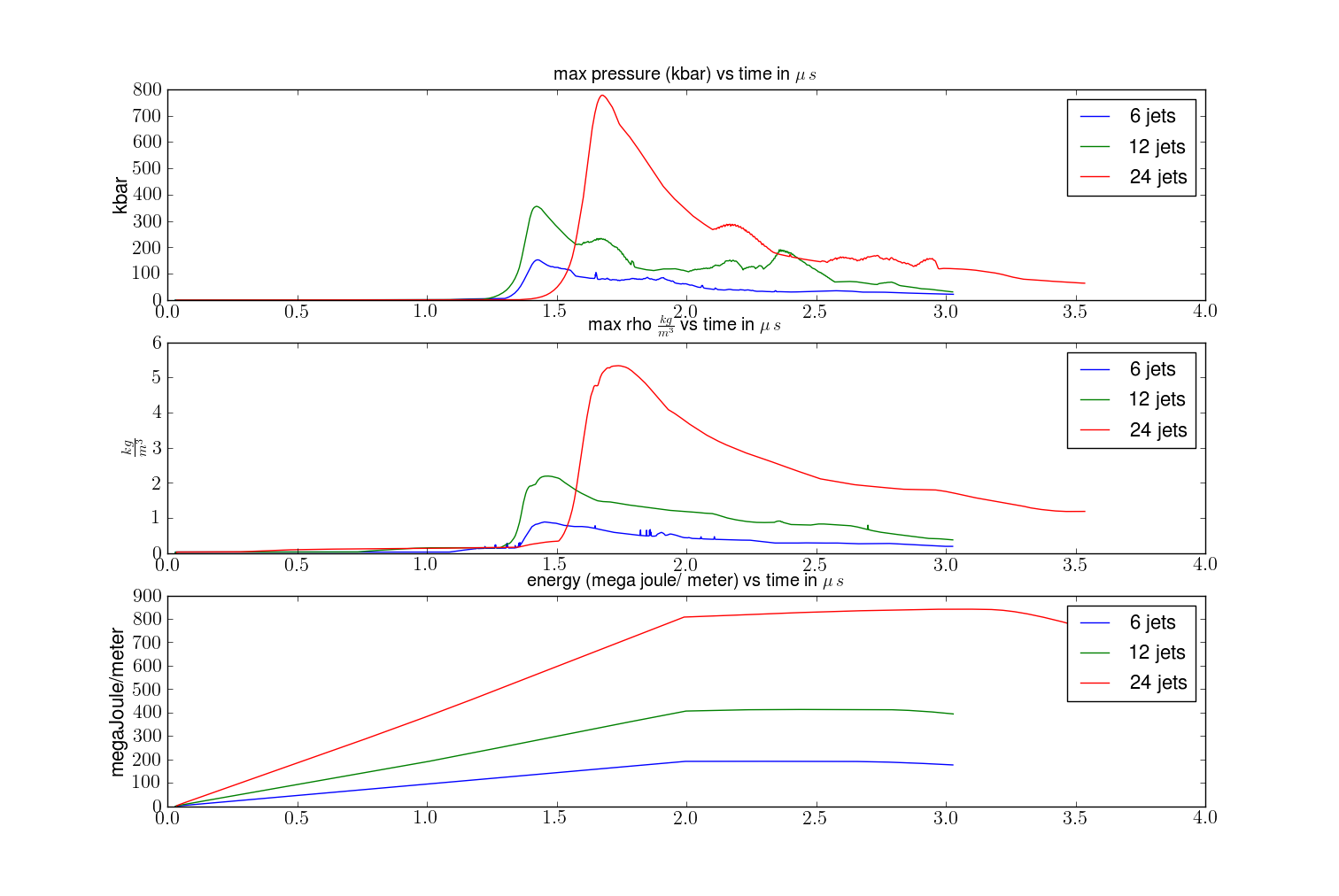}
	\end{center}\caption{Merging Mach 80 gas dynamic jets without radiation.  Peak pressure reaches $800\text{kbar}$ in the 24 jet simulation.}\label{F:eulerHist80}
\end{figure}

Figure \ref{F:eulerHist40} and figure \ref{F:eulerHist80} show peak pressure, density and the total energy in the system for merging neutral jets at Mach 40 and 80 using 6, 12 and 24 jets.  No radiation was included in these simulations.  Peak pressure reaches $140 \text{kbar}$ and density reaches $3 kg/m^{3}$ in the Mach 40 case and $800\text{kbar}$ and $5.5 kg/m^{3}$ in the Mach 80 case.  Total energy of the system is also plotted.  Jets enter the domain through the boundary so total energy increases until the jets are turned off at which point the energy remains constant.  As expected, pressure and density peak and then rapidly decline.  Peak pressure is increased as the number of jets is increased in agreement with work performed by Cassibry \cite{Cassibry2009}.

\begin{figure}
	\begin{center}
		\includegraphics[scale=0.4]{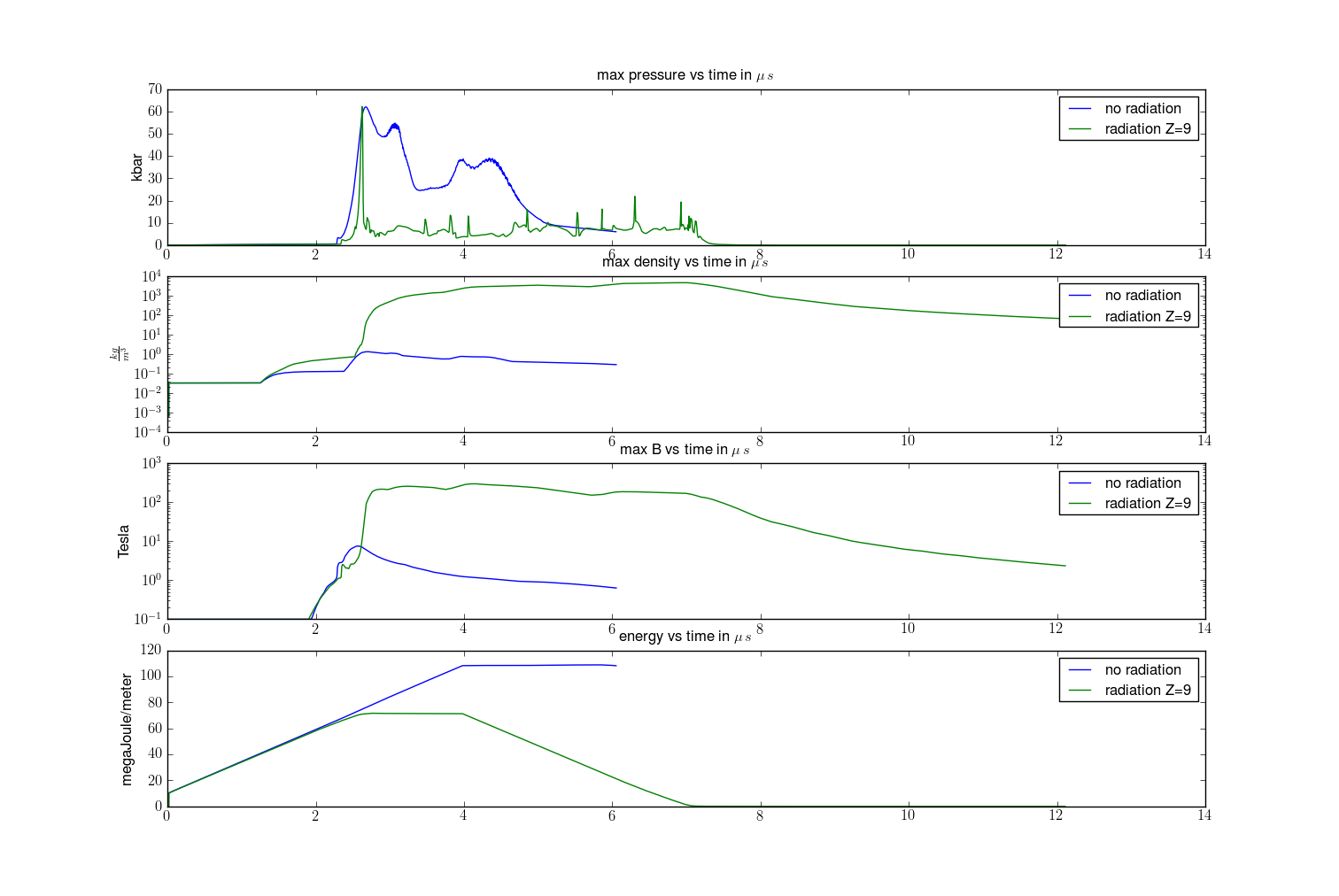}
	\end{center}\caption{Mach 40 MHD jet merging simulations with and without using optically thin bremsstrahlung radiation model.  Log plots of the peak magnetic field show that the peak field reaches about 100 times the value achieved in the non-radiating case.}\label{F:mhdHist40}
\end{figure}

\begin{figure}
	\begin{center}
		\includegraphics[scale=0.4]{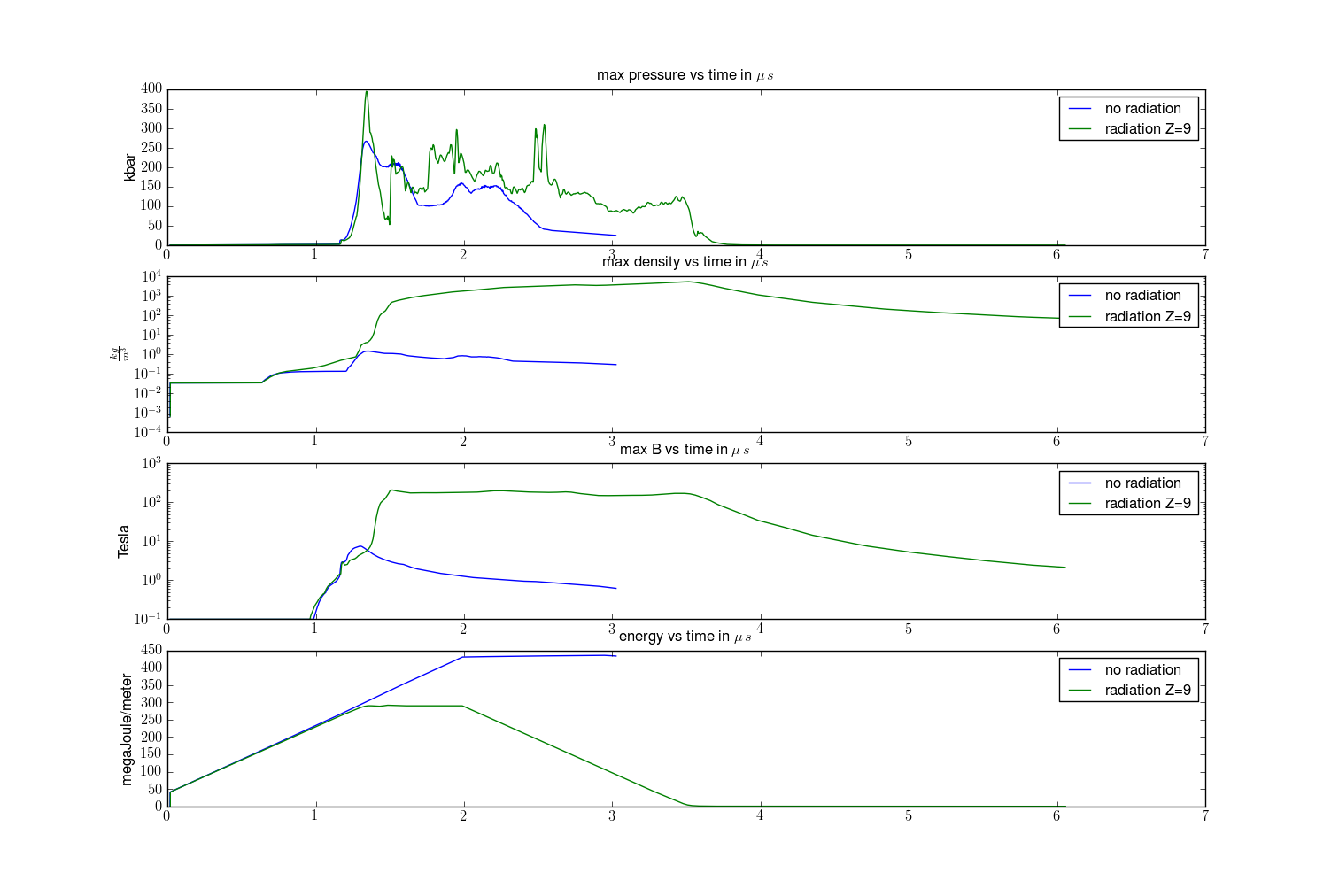}
	\end{center}\caption{Mach 80 MHD jet merging simulations with and without using optically thin bremsstrahlung radiation model.  The field compresses to about 100 times the non-radiating case, similar to what occurs for the Mach 40 jets.}\label{F:mhdHist80}
\end{figure}

Figure \ref{F:mhdHist40} and \ref{F:mhdHist80} show plots of peak $B_{z}$  density, pressure and total system energy of Mach 40 and Mach 80 MHD jets with target a $0.1\text{Tesla}$ initial
magnetic field.  The initial target configuration is not in equilibrium and would expand if it were not for compression by the plasma jets.  The field can be though of as an idealized solenoidal field.  Results are shown with and without using the optically thin radiation model.  Note that in \ref{F:mhdHist40} and \ref{F:mhdHist80} the magnetic field and density have been plotted on a log plot.  When radiation is included in the model the peak magnetic field reaches about 100 times its value in the non-radiating case.  What's more, peak field and density are maintained constant for several microseconds, corresponding to the length of time the jets are turned on.  The results suggest that plasma jets might be able to be used used to obtain long lasting strong magnetic fields, furthermore they provide an upper limit on what to expect when complete radiation models are included (including reabsorption).

\begin{figure}
	\begin{center}
		\includegraphics[scale=0.4]{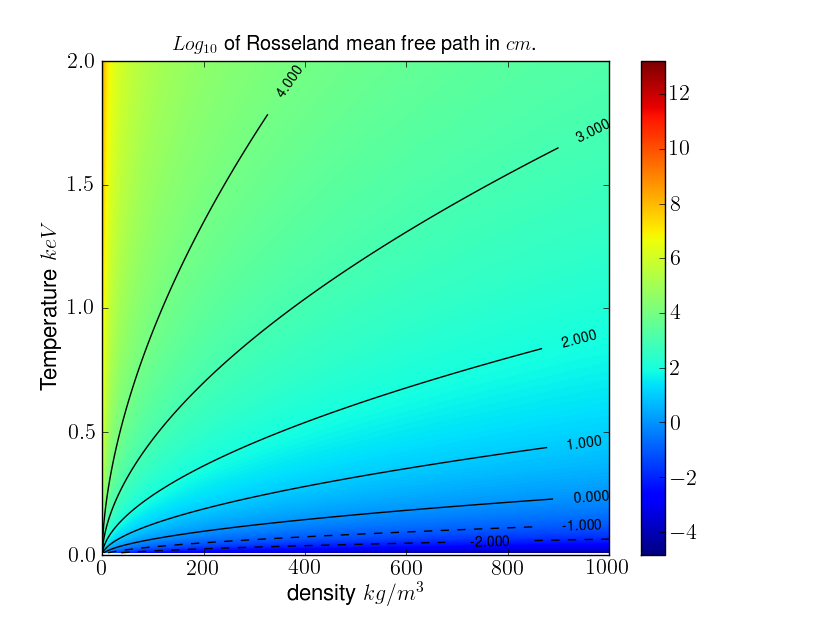}
	\end{center}\caption{Plot of bremsstrahlung Rosseland mean free path vs temperature and density assuming constant $Z_{eff}=9$.  In much of the plasma the optically
	thin model is accurate, however during liner formation the plasma will radiate significantly cooling it down at high density and decreasing the bremsstrahlung mean free
	path until a significant part of the radiation is re-absorbed.  A radiation diffusion approximation can be used to capture this effect.  This also indicates the importance of determining $Z_{eff}$ accurately, as a larger $Z_{eff}$ will decrease the mean free path and a smaller $Z_{eff}$ will increase the mean free path by a significant amount.}\label{F:absorption}
\end{figure}

Using the Rosseland mean free path for bremsstrahlung radiation
\begin{equation}
	L\approx \frac{64.0 A^{2}\,T_{e}^{7/2}}{Z_{eff}^{3}\,\rho^{3}}\,\text{cm}
\end{equation}
 (defined in \cite{Atzeni1987} for example) where $A$ is the atomic mass number, $T_{e}$ is the electron temperature in keV and $\rho$ is the mass density in $\frac{g}{cm^{3}}$, we can 
 arrive at an estimate of the importance of absorption in different regimes.  In these simulations peak plasma densities of 
 around $5000\frac{\text{kg}}{m^{3}}$ were achieved while the initial jet was about $.03\frac{\text{kg}}{m^{3}}$.  Figure \ref{F:absorption} shows
 the Rosseland mean free path (in centimeters) for bremsstrahlung radiation as a function of density and temperature.  When the mean free path
 is below about $1 \text{cm}$ reabsorption will be significant.  The plot suggests that reabsorption will be significant in the jet plasma during merging
 after it radiates a significant amount of energy away bringing it to around 100eV in temperature.  This plot assumes a constant $Z_{eff} = 9$.  This also suggests
 that radiation from the magnetized target will be partially absorbed by the dense relatively cold liner.
\newpage

\section{Conclusion}
First steps in modeling plasma jet merging have been performed in 2D using ideal gas dynamic and ideal MHD equations with a simple optically thin radiation model assuming constant $Z_{eff}$.  Gas-dynamic modeling without radiation or targets showed that merging jet peak pressures on the order of 100s of kBar could be achieved for the conditions simulated, peak pressure increasing with the number of jets.  Results indicate late time Rayleigh Taylor instabilities in all cases and the heavy influence of radiation in the MHD simulations.  Including radiation results in field enhancement 100 times that observed in simulations without radiation and resulting peak field compression of several microseconds, substantially different than the non-radiating case.  The radiation simulations illustrate the importance of properly capturing the argon plasma ionization state in addition to the importance of modeling the radiation properly.  The optically thin model becomes invalid as the liner density rises and we would ultimately expect lower peak field compression as a result.  The radiation in the jet effectively increases the jet Mach number, therefore increasing the ram efficiency of the jet.  In these simulations, both jet and target were argon plasmas. In a real device, the target plasma would be deuterium, while the jets argon or some other high Z gas resulting in large radiation losses in the jet and lower radiation losses in the target. An accurate study will require complete atomic physics requiring multi-species MHD and radiation in both the optically thin and diffusion limit.  These efforts will be part of future investigations.


\begin{thebibliography}{5}
\bibitem{facets}
John R. Cary, Ammar Hakim, Mahmood Miah, Scott Kruger, Alexander
Pletzer, Svetlana Shasharina, Srinath Vadlamani, Alexei Pankin, Ronald
Cohen, Tom Epperly, Tom Rognlien, Richard Groebner, Satish Balay, Lois
McInnes, Hong Zhang, ÒFACETS Ð a Framework for Parallel Coupling of
Fusion ComponentsÓ, The 18th Euromicro International Conference on
Parallel, Distributed and Network-Based Computing. Pisa, Italy. 2010.


\bibitem{Thio1999}
Y.C. F. Thio et al, \text{Magnetized Target Fusion in a Spheroidal Geometry with Standoff Drivers},  \emph{Current Trends in International Fusion Research}, 1999,  p. 113

\bibitem{Thio2001}
Y.C. F. Thio et al, \text{A Physics Exploratory Experiment on Plasma Liner Formation},  \emph{Journal of Fusion Energy},  volume 20, 2001,  1--11

\bibitem{Hsu2008}
Scott C. Hsu,\text{Technical Summary of the First U.S. Plasma Jet Workshop} \emph{Journal of Fusion Energy}
volume 75, number 10, 1995, 246--257

\bibitem{Lindemuth1995}
I.R. Lindemuth et al, \text{Target plasma formation for magnetic compression/magnetized target fusion} \emph{Physical Review Letters}
volume 75, number 10, 1995, 1953--1956

\bibitem{Parks2008}
P.B. Parks, \text{On the efficacy of imploding plasma liners for magnetized fusion target compression}, \emph{Physics of Plasmas}
15, 2008, 062506--062512

\bibitem{Atzeni1987}
S. Atzeni, \text{The physical basis for numerical fluid simulations in laser fusion}, \emph{Plasma Physics and Controlled Fusion}
Volume 29, Number 11, 1987, 1535--1604

\bibitem{Degnan1999}
J.H. Degnan et al, \text{Compression of Plasma to Megabar Range using Imploding Liner}, \emph{Physical Review Letters}
Volume 82, Number 13, 1999, 2681--2684

\bibitem{Cassibry2008}
Jason Cassibry, \text{Modeling of formation and implosion of plasma liners by discrete jets}, \emph{39th AIAA Plasmadynamics and Lasers Conference}, (2008)

\bibitem{Cassibry2009}
Jason Cassibry et al, \text{Hydrodynamic Modeling of the Plasma Liner Experiment}, \emph{51st Annual APS-DPP Meeting}, (2009)

\bibitem{Waagan2009}
K. Waagan, \text{A positive MUSCL-Hancock scheme for ideal magnetohydrodynamics}, \emph{Journal of Computational Physics}, 228 (2009), 8609--8626

\bibitem{Stone2009}
James M. Stone et al, \text{A simple unsplit Godunov method for multidimensional MHD}, \emph{New Astronomy}, 14 (2009), 139--148

\bibitem{Dedner2002}
A. Dedner et al, \text{Hyperbolic Divergence Cleaning for the MHD Equations}, \emph{Journal of Computational Physics}, 175 (2002), 645--673

\end{thebibliography}
\end{document}